\documentclass[useAMS,usenatbib,usegraphicx]{mn2e}

\newcommand{\etal}{et al.\ }
\newcommand{\etalb}{et al.}
\newcommand{\beq}{\begin{equation}}
\newcommand{\beqa}{\begin{eqnarray}}
\newcommand{\eeq}{\end{equation}}
\newcommand{\eeqa}{\end{eqnarray}}

\title[Light-cone anisotropy in 21cm fluctuations during the epoch of
reionization] {Light-cone anisotropy in 21cm fluctuations during the epoch
of reionization}

\author[R. Barkana and A. Loeb]{Rennan Barkana$^{1}$ and Abraham Loeb$^{2}$
\thanks{E-mail: barkana@wise.tau.ac.il (RB); aloeb@cfa.harvard.edu (AL)}\\
$^{1}$School of Physics and Astronomy, The Raymond and Beverly Sackler
Faculty of Exact Sciences,\\ Tel Aviv University, Tel Aviv 69978,
ISRAEL\\ $^{2}$Astronomy Department, Harvard University, 60 Garden
Street, Cambridge, MA 02138, USA}

\begin{document}

\pagerange{\pageref{firstpage}--\pageref{lastpage}} \pubyear{2005}

\maketitle

\label{firstpage}

\begin{abstract}
The delay in light travel time along the line of sight generates an
anisotropy in the power spectrum of 21cm brightness fluctuations from
the epoch of reionization. We show that when the fluctuations in the
neutral hydrogen fraction become non-linear at the later stages of
reionization, the light-cone anisotropy becomes of order unity on
scales $\ga 50$ comoving Mpc. During this period the density
fluctuations and the associated anisotropy generated by peculiar
velocities are negligible in comparison.
\end{abstract}


\section{Introduction}

Fluctuations in the 21cm brightness from cosmic hydrogen at redshifts
$z\ga 6$ were sourced by the primordial density perturbations from
inflation \citep{loeb04,Bar05a,Bar05c} as well as by the radiation
from galaxies \citep{Scott,Madau,Fur04,Bar05b}. These two different
components can be separated based on the angular dependence of the
21cm fluctuation power spectrum which is induced by peculiar
velocities \citep{Bha,Bar05a}. Uncertainties in the cosmological
parameters lead to an additional apparent anisotropy due to the
Alcock-Paczy\'{n}ski effect; after accounting for constraints from the
cosmic microwave background, this effect can still produce up to a
$10\%$ anisotropy which may be identifiable because of its particular
angular structure \citep{nusserAP,barkanaAP}. Several low-frequency
arrays that could potentially detect the redshifted 21cm signal are
being built around the globe, including the {\it Primeval Structure
Telescope} (web.phys.cmu.edu/$\sim$past), the {\it Mileura Widefield
Array} (web.haystack.mit.edu/arrays/MWA), and the {\it Low Frequency
Array} (http://www.lofar.org).

In this work we examine an additional source of anisotropy in the 21cm
power spectrum between the directions parallel and transverse to the
line of sight. Previous discussions ignored the delay in light travel
time (i.e., the ``light-cone'' constraint) along the line of sight
[but see a rough estimate of the effect at the beginning of
reionization in an appendix in \citet{mcquinn}]. In particular,
although the 21cm power spectrum is expected to be measured through
averaging over three-dimensional volumes of finite radial extent, the
power spectrum was previously evaluated at a fixed time slice of the
universe, ignoring the fact that a fixed observing time implies a
varying emission time as a function of distance from the
observer. Here we evaluate the amplitude of the anisotropy which is
sourced by this delay. This time-delay effect is not related to real
causal effects or light-crossing times. Since two points at a given
separation are in general seen at different redshifts, the correlation
function, averaged over all such points, is affected by the change
with time of the statistics of ionization (i.e., the distribution of
H~II regions and their correlation with the underlying density field).

\section{Model}

\citet{BLflucts} showed that the biased large-scale fluctuations in
the number density of galaxies at high redshift imply that the
characteristic bubble radius is determined by correlated groups of
galaxies rather than by the bubble sizes of individual
galaxies. \citet{Fur04} developed a semi-analytic model based on this
realization, and used it to predict the distribution of H~II bubble
sizes around a given point (i.e., the one-point distribution of
bubbles). Unfortunately, this model cannot be directly applied to
estimating two-point distributions. Upcoming 21cm experiments will
only have the sensitivity to probe statistical measures of the 21cm
fluctuations and will therefore focus on the three-dimensional power
spectrum (or, equivalently, the two-point correlation function). We
formulate here an approximate model of the 21cm correlation function
in order to estimate the time-delay anisotropy.

Suppose that gas with an overdensity $\delta$ and a neutral fraction
$x_n$ lies in some direction at a distance corresponding to 21cm
absorption at a redshift $z$. Then the resulting 21cm brightness
temperature offset relative to the CMB is \citep{Madau} \beq T_b =
28.8\, (1+\delta) x_n\, \left( \frac{\Omega_b h} {.033} \right)
\left( \frac{\Omega_m} {.27} \right)^{- \frac{1} {2}}
\left({{1+z}\over{10}}\right)^{1 \over 2}\, {\rm mK}\ , \label{eq:Tb} \eeq
where we have assumed that the IGM has been heated so that $T_s \gg
T_{\rm CMB}$, and we have substituted the concordance values
\citep{CMB} for the cosmological parameters $\Omega_b$, $h$, and
$\Omega_m$. In order to calculate the two-point correlation function
of the 21cm brightness temperature, we must construct the correlated
distribution of $\delta$ and $x_n$ at two different points
observed at two possibly different redshifts. We first estimate the
characteristic bubble radius $R_{\rm ch}$ at each redshift, and
calculate the mean ionized fraction of the interior of regions of
radius $R_{\rm ch}$ as a function of their mean densities. We then
calculate the correlation function at two points by integrating over
the joint probability distribution of their overdensities.

We first must estimate the characteristic radius $R_{\rm ch}$ at a
redshift $z$, defined so that for a given point $P$, the average
number of ionizing sources within $R_{\rm ch}$ typically determines
whether $P$ is ionized or not. In estimating $R_{\rm ch}$ we must
balance two competing trends. On the one hand, if a large region
contains enough sources to fully reionize itself, then it will indeed
fully reionize regardless of whether small portions within it contain
a low density of galaxies; this simple fact favors larger radii over
smaller ones. On the other hand, the $\Lambda$CDM density power
spectrum implies that the typical magnitude of density fluctuations
decreases with scale, so that large regions are unlikely to have a
strong positive fluctuation that will increase the number of galaxies
sufficiently for them to fully reionize. Thus, we estimate $R_{\rm
ch}$ as the largest radius for which a 1-$\sigma$ density fluctuation
results in full reionization of a sphere of this radius.

Towards the end of reionization, this simple estimate of $R_{\rm ch}$
approaches infinity, but \citet{WL04} showed that the bubble radius is
limited by the finite light travel time to $\sim 70$ comoving Mpc at
redshift $\sim 6$; we generalize their argument and set an upper limit
on $R_{\rm ch}$ at any stage of reionization. In general, in order to
have a coherent ionized bubble of radius $R_{\rm ch}$, various
portions of the bubble, some of which are relatively rich with
galaxies and some of which have fewer ionizing sources, must
effectively exchange photons between them. There are two important
timescales: the typical (i.e., 1-$\sigma$) scatter $t_{\rm scatter}$
in the time when different regions of size $R_{\rm ch}$ reach a given
stage of galaxy formation (in terms of a given number density of
galaxies); and the light crossing time $t_{\rm light}$ of the radius
$R_{\rm ch}$. Now, if $t_{\rm light} > t_{\rm scatter}$ at a given
redshift $z$, then such a large region will not constitute a single,
coherent bubble; e.g., a low-density portion of this region that has
not produced enough photons to reionize itself will indeed not be
fully reionized at redshift $z$, since the extra ionizing photons
needed from the distant portions of the region will typically have
been produced a time $\sim t_{\rm scatter}$ earlier but require a time
$\sim t_{\rm light}$ to reach the low-density region. We follow
\citet{WL04} in using the extended Press-Schechter model
\citep{bond91} to estimate this upper limit on $R_{\rm ch}$.

The progress of reionization depends on the total collapse fraction of
gas in galactic halos, $F_{\rm col}$, which is determined by
integrating the halo mass function over all halos that host
galaxies. We use the halo mass function of \citet{shetht99} which fits
numerical simulations accurately \citep{jenkins}. In order to
calculate fluctuations in the gas fraction that resides in galaxies
within different regions, we adjust the mean halo distribution based
on the prescription of \citet{BLflucts} which fits a broad range of
simulations. As gas falls into a dark matter halo, we assume it can
fragment into stars only if its virial temperature is above $10^4$K,
which allows for efficient cooling mediated by atomic transitions. We
assume that stars dominate over mini-quasars, as expected at high
redshift \citep{WL05} [but see \citet{quasars}]. The stellar spectrum
depends on the initial mass function (IMF) and the metallicity of the
stars. We consider two examples. The first is the locally-measured IMF
of \citet{scalo} with a metallicity of $1/20$ of the solar value,
which we refer to as Pop II stars. The second case, labeled as Pop III
stars, consists entirely of zero metallicity $M \ga 100\, M_{\odot}$
stars, as expected for the earliest galaxies \citep{Bromm04}.

If we were to treat each sphere of radius $R_{\rm ch}$ as an isolated
region, then the mean ionized fraction in a region of mean
perturbation $\bar{\delta}$ would simply be related to the gas
fraction in galaxies $F_{\rm col}$: \beq \bar{x}_i = \left(1+
\frac{N_{\rm ion}} {0.76} \right) F_{\rm col}(z,R_{\rm ch},
\bar{\delta})\ , \eeq where $N_{\rm ion}$ is the overall efficiency of
emission of ionizing photons per baryon in galactic halos, the factor
of 0.76 converts between hydrogen and baryon number densities, and the
(minor) addition of 1 counts the gas in galaxies (which need not be
ionized by the photons that escape from galaxies). Note that $N_{\rm
ion}$ equals the number of ionizing photons produced per baryon in
stars, multiplied by the star formation efficiency $f_*$ and by the
escape fraction $f_{\rm esc}$ of ionizing photons from halos.

In reality different regions are not independent. The overdense
regions produce more ionizing photons than they need to fully reionize
themselves, and the extra photons expand into the surrounding
lower-density regions. In order to make our model consistent with the
total number of ionizing photons produced at each redshift, we
increase the $\bar{x}_i$ values by the amount necessary to achieve
consistency. Now, the leakage of additional photons is expected to
depend on the mean $\delta$ of the target region. In particular, the
deepest voids are likely to be the farthest distance away from the
densest regions. We estimate the relative fractions distributed into
different regions as follows. For each $\bar{\delta}$, we assume that
the increase in $\bar{x}_i$ within a region of that $\bar{\delta}$ is
proportional to the chance that such a sphere lies next to a sphere
with $\bar{\delta}=\delta_i$, where $\delta_i$ is a high overdensity
representative of those regions that have overproduced ionizing
photons. We adopt for $\delta_i$ the minimum value of $\bar{\delta}$
needed for a sphere to achieve an $\bar{x}_i=1$ on its own, i.e.,
without the help of surrounding regions. In general, we emphasize that
the redistribution of photons has a modest effect within our model,
consistent with our selection of $R_{\rm ch}$ in such a way that
photons from larger distances are unlikely to play a major role in the
statistics of reionization.

When we calculate the correlated ionization states of two points at
redshifts $z_1$ and $z_2$, we assume that the ionization probability
of each point is determined by the perturbations $\bar{\delta}$ of the
surrounding regions of radius $R_{\rm ch}(z_1)$ and $R_{\rm ch}(z_2)$,
respectively. However, the 21cm brightness temperature $T_b$ is
determined not only by the neutral fraction but also by the
perturbation $\delta$, which is in general different from the
spherically-averaged perturbation $\bar{\delta}$. While $\bar{\delta}$
is determined by large-scale power above the scale of $R_{\rm ch}$,
the difference $\delta - \bar{\delta}$ is determined by the additional
small-scale power within the bubble. In the spirit of the extended
Press-Schechter model \citep{bond91}, we assume that $\bar{\delta}$
and $\delta - \bar{\delta}$ are approximately statistically
independent. We thus add the portion $\delta - \bar{\delta}$ at each
point independently of the values of $\bar{\delta}$ and the neutral
fraction $\bar{x}_{\rm n}$ of the two regions, with a correlation
function for the small-scale portion of $\delta$ set equal to the
total $\xi(r)$ of the density field minus the smoothed $\xi(r)$ that
determines the joint statistics of $\bar{\delta}$.

When calculating the correlated ionization states of two points we
must carefully interpret the values of $\bar{x}_{n,1}$ and
$\bar{x}_{n,2}$ in the regions surrounding the two points as {\it
probabilities}\/ of having neutral gas. For example, the mean value of
$x_n^2$ at point 1 is not $\bar{x}_{n,1}^2$, which it would be if the
gas were partially ionized with a neutral fraction of $\bar{x}_{n,1}$,
but is instead $\bar{x}_{n,1}$ since the gas has $x_n=1$ with
probability $\bar{x}_{n,1}$ and $x_n=0$ with probability
$1-\bar{x}_{n,1}$. Thus, the mean expected value of $x_{n,1} x_{n,2}$
must be calculated as the joint probability that both points are
neutral. Given the individual probabilities $\bar{x}_{n,1}$ and
$\bar{x}_{n,2}$, there is some probability $p_{\rm same}$ that the two
points are ionized within the same bubble. Assuming that their
ionization probabilities are independent if they are not within the
same bubble, we find that \beq \langle x_{n,1} x_{n,2} \rangle =
\bar{x}_{n,1} \bar{x}_{n,2} / (1-p_{\rm same})\ . \eeq Now, in order
to estimate $p_{\rm same}$ for two points separated by a comoving
distance $r$, we note that if one point is ionized by a bubble of
radius $R_{\rm ch}$ then the probability that this (randomly-placed)
bubble contains a second point a distance $r$ away is zero if $r> 2
R_{\rm ch}$ and \beq f(r,R_{\rm ch}) = 1-\frac{3 r} {4 R_{\rm ch}^3}
\left( R_{\rm ch}^2 - \frac{1}{12} r^2 \right) \eeq otherwise
\citep{Fur04}. This yields two estimates of $p_{\rm same}$, starting
from either point, so we take the average as our best estimate: \beq
p_{\rm same} \approx \frac{1}{2} \left[ ( 1-x_{n,1}) f(r,R_{\rm
ch}(z_1)) + ( 1-x_{n,2}) f(r,R_{\rm ch}(z_2)) \right]\ , \eeq with the
constraint that $p_{\rm same}$ must not be higher than the separate
ionization probabilities of each of the two points.

The two-point correlation function $\xi$ of the 21cm brightness
temperature $T_b$ is a function of the distance $r$ between the two
points and their two redshifts. In order to focus on the anisotropy as
a function of the angle $\theta$ between the line of sight and the
vector connecting the two points, we parameterize $\xi$ as a function
of $r$, $\mu \equiv {\rm cos} \theta$, and the redshift $z$ at the
midpoint (in terms of comoving distance) of the two points. We
calculate the expectation value \beq \xi =
\left \langle \left( T_{b,1} - \bar{T}_b(z_1) \right) \times \left(
T_{b,2} - \bar{T}_b(z_2) \right) \right \rangle\ , \eeq assuming that
the mean $\bar{T}_b$ at the appropriate redshift has been subtracted
from $T_b$ at each point. Such a subtraction is expected in future
observations since the radio measurements take the Fourier transform
of the intensity in the $x$ and $y$ directions, removing any
sensitivity to a varying $\bar{T}_b(z)$ as long as the line
$k_x=k_y=0$ in Fourier space is avoided. We also assume in our results
below that $T_b(z_1)$ has been multiplied by $[(1+z)/(1+z_1)]^{1/2}$,
and similarly for $T_b(z_2)$, in order to avoid any anisotropy due to
the simple redshift scaling in equation~(\ref{eq:Tb}).

We calculate the correlation function explicitly by averaging over the
joint normal distribution of the perturbations $\bar{\delta}$ of the
regions of radii $R_{\rm ch}(z_1)$ and $R_{\rm ch}(z_2)$ surrounding
the two points. We also calculate the mean value $\bar{T}_b(z)$ at
each redshift by averaging the quantity $[(1+\delta) x_n]$ over the
probability distribution. Note that we do not include peculiar
velocity fluctuations here since we focus on the late stages of
reionization and show that in this regime the 21cm fluctuations and
their anisotropy are dominated by ionization fluctuations.

\section{Results}

We illustrate our predictions for two cases which both assume atomic
cooling, a star formation efficiency $f_*=2\%$, and an escape fraction
of ionizing photons from halos of $f_{\rm esc}=10\%$; however, our
low-$z$ case assumes Pop II stars (\S~2) and achieves complete
reionization at redshift 7.35, while our high-$z$ case assumes Pop III
stars and completes reionization already at redshift 13.7. Note that
the parameter values given here assume that only one ionizing photon
per hydrogen atom is needed in order to achieve reionization of the
IGM. If we assumed instead that several photons are needed due to
recombinations, then this would require an increase in the value of
$f_*$ by the same factor as the required number of photons per
atom. This approximation is fairly accurate as long as the number of
times each hydrogen atom must be reionized is roughly the same in
different regions. Indeed, the recombination rate is not expected to
vary greatly in different large-scale regions, since the fluctuation
in the mean density of such regions is small. Recombinations can be
included in a future refinement of our model.

\begin{figure}
\includegraphics[width=84mm]{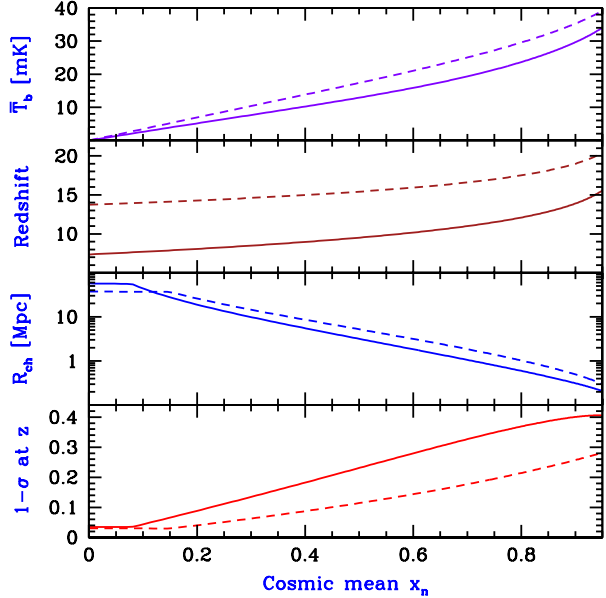}
\caption{Reionization history. We show the cosmic mean 21cm brightness 
  temperature (top panel), redshift $z$ (next lower panel), the
  comoving characteristic radius $R_{\rm ch}$ (next lower panel), and
  the 1-$\sigma$ density fluctuation in spheres of radius $R_{\rm ch}$
  at $z$ (bottom panel), each vs.\ the cosmic mean neutral fraction
  $x_n$. Each panel shows our low-$z$ case (solid curve) and our
  high-$z$ case (dashed curve).}
\label{fig:zOfxe}
\end{figure}

Figure~\ref{fig:zOfxe} shows the two reionization histories, as well
as the characteristic radius $R_{\rm ch}$ and the standard deviation
of the mean density $\bar{\delta}$ in a sphere of radius $R_{\rm
ch}$. At each redshift the probability distribution of $\bar{\delta}$
in the spherical region is a Gaussian with this standard deviation; we
have not included non-linear corrections which are particularly
important early in reionization when $R_{\rm ch}$ is relatively
small. The characteristic bubble radius increases during the epoch of
reionization as larger groups of galaxies are able to fully reionize
the region containing them. At the later stages of reionization, when
the cosmic mean $x_n \la 10\%$, $R_{\rm ch}$ is determined by the
light travel time of ionizing photons.
   
\begin{figure}
  \includegraphics[width=84mm]{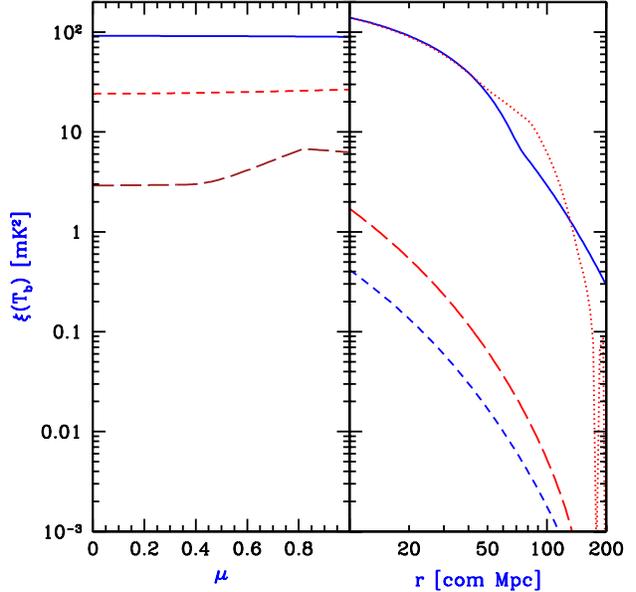}
\caption{Correlation function of 21cm brightness temperature as a
  function of $r$ or $\mu$, at a late stage of reionization
  ($\bar{x}_n=0.25$) by Pop III stars. The left panel considers
  comoving $r=20$ Mpc (solid curve), 50 Mpc (short-dashed curve), and
  100 Mpc (long-dashed curve). The right panel considers angle $\mu=0$
  (solid curve) and $\mu=1$ (dotted curve). Actually shown is $|\xi|$,
  since $\xi$ becomes negative when $r \sim 200$ Mpc and $\mu=1$. Also
  shown are the predictions of a model with density and peculiar
  velocity fluctuations but no ionization fluctuations, for $\mu=0$
  (short-dashed curves) and $\mu=1$ (long-dashed curves).}
\label{fig:xiOfr1}
\end{figure}

In Figure~\ref{fig:xiOfr1} we illustrate the correlation function
$\xi$ of 21cm brightness temperature late in reionization in the
high-$z$ case. If we considered density fluctuations alone (along with
a uniform $x_n(z)$ equal to its cosmic mean value at each redshift),
then $\xi(r)$ would roughly equal the 1-$\sigma$ density fluctuation
on scale $r$, times the cosmic mean $T_b$, all squared.  The growth
factor and $x_n(z)$ both evolve slowly at this stage of reionization
and produce a negligible anisotropy. Adding in peculiar velocities
generates a significant anisotropy, but only causes fluctuations that
are of order the density fluctuations. However, the ionization
fluctuations are far larger and indeed non-linear, since while the
bubble radius $R_{\rm ch}$ is large and the corresponding density
fluctuations are small (see Figure~\ref{fig:zOfxe}), various regions
can have values of $x_n$ covering almost the entire possible range of
0--1. In particular, we find that although the 1-$\sigma$ density
fluctuation on the scale of $R_{\rm ch} \sim 20$ Mpc is $\sim 0.05$ at
this redshift, a negative 2-$\sigma$ void on this scale has an $x_n
\sim 60\%$ (compared to a cosmic mean $x_n$ of $25\%$), while a
positive 1-$\sigma$ overdense region is almost completely reionized.

\begin{figure}
\includegraphics[width=84mm]{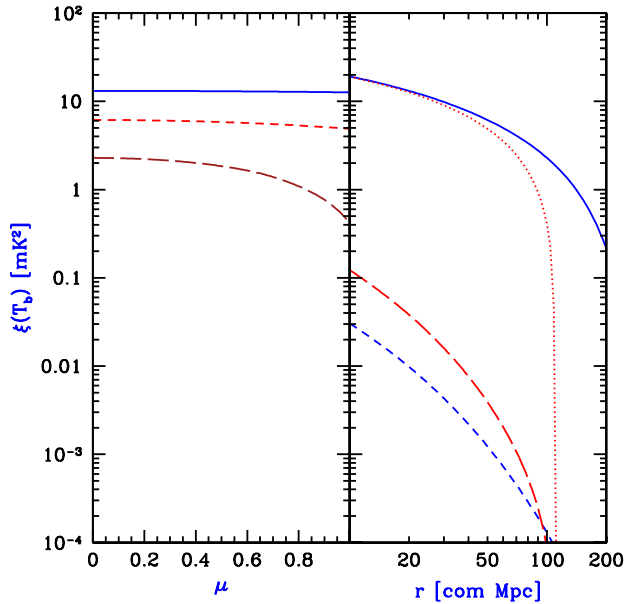}
\caption{Correlation function of 21cm brightness temperature, same as
  Figure~\ref{fig:xiOfr1}, except near the end of reionization
  ($\bar{x}_n=0.05$) by Pop II stars}
\label{fig:xiOfr2}
\end{figure}

Figure~\ref{fig:xiOfr2} illustrates the 21cm correlation function near
the end of reionization in the low-$z$ case. The ionization
fluctuations again dominate, since the deep voids still have a
substantial neutral fraction that is much higher than the cosmic mean
$x_n$. The strong anisotropy here is related to the rapid reionization
of the deep voids at the end of reionization; in particular, $\xi$
drops to zero when the lower-redshift point approaches the redshift
where cosmic reionization is completed. Although we have predicted the
general trends, we emphasize that the results at the end of
reionization depend on how rapidly the deep voids get reionized by
photons from denser regions, and thus precise quantitative predictions
must await fully self-consistent calculations of large-scale radiative
transfer.

\begin{figure}
  \includegraphics[width=84mm]{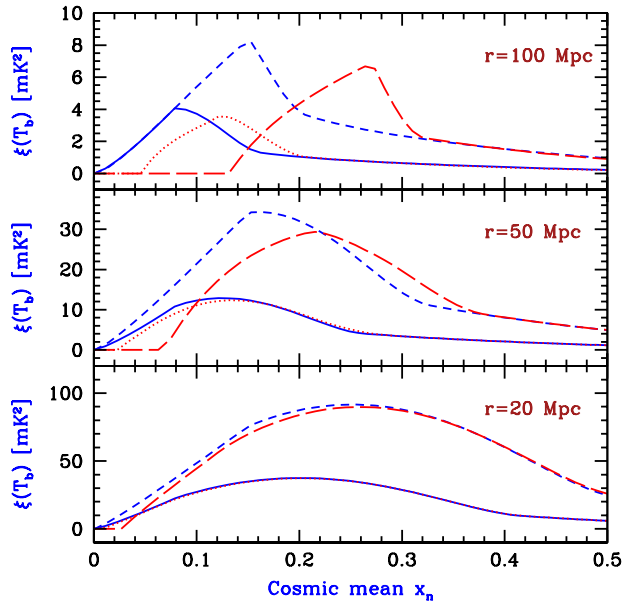}
\caption{Correlation function of 21cm brightness temperature as a
  function of the cosmic mean neutral fraction $x_n$. Each panel
  corresponds to a comoving distance $r$ as indicated. We consider Pop
  II stars and an angle of $\mu =0$ (solid curves) or 1 (dotted
  curves), or Pop III stars and an angle of $\mu =0$ (short-dashed
  curves) or 1 (long-dashed curves).}
\label{fig:xiOfz}
\end{figure}

Figure~\ref{fig:xiOfz} summarizes the evolution of the 21cm
correlation function during the latter half of the reionization
epoch. Early in reionization, ionization affects only the rare
high-density regions, and the anticorrelation of $\bar{x}_n$ and
$\bar{\delta}$ leads to a low $\xi$ even though the cosmic mean
brightness temperature is relatively high. Later on, as reionization
spreads to more typical regions, $\bar{\delta}$ fluctuations produce
amplified $\bar{x}_n$ fluctuations and $\xi$ increases, only to
decrease at the end of reionization when the cosmic mean $x_n$
approaches zero. The anisotropy is most easily observable during the
late stages where the signal is above 1 mK and is highly
anisotropic. Note that the anisotropy is stronger in the case of Pop
III stars, since in that case reionization occurs at a higher redshift
and is caused by rarer halos, whose number density changes more
rapidly with redshift during reionization.

\section{Conclusions}

At the later stages of reionization, ionization fluctuations dominate
the 21cm power spectrum (Figures~\ref{fig:xiOfr1} and
\ref{fig:xiOfr2}) and the light travel-time delay generates a strong
anisotropy. The signal is around a few mK on scales and redshifts
where the time-delay anisotropy is large (Figure~\ref{fig:xiOfz}),
while noise and foreground-subtraction should still allow even
first-generation 21-cm experiments to reach a sensitivity $\sim 1$ mK
on large scales \citep{miguel,mcquinn}.

The strong line-of-sight anisotropy in the ionization fluctuations
arises in our model since the cosmic mean neutral fraction changes
rapidly with redshift and its fluctuations are highly non-linear. The
scale and angular dependence of the light-cone anisotropy must be
modelled carefully in order to separate the inflationary initial
conditions from astrophysical effects
\citep{Bar05a}. Large-scale numerical simulations of reionization
\citep{Iliev05} could provide guidance for this difficult task, but
fully self-consistent simulations with hydrodynamics and radiative
transfer are required. Alternatively, the separation of the
``physics'' (i.e., the inflationary initial conditions) from the
``astrophysics'' may require observations at higher redshifts, when
the neutral fraction is closer to unity and the 21cm power spectrum is
dominated by density and peculiar velocity fluctuations.

\section*{Acknowledgments}

R.B. is grateful for the kind hospitality of the {\it Institute for Theory
\& Computation (ITC)} at the Harvard-Smithsonian CfA where this work
began, and acknowledges support by Harvard university and ISF grant
629/05. This work was supported in part by NASA grants NAG 5-13292,
NNG05GH54G, and NSF grant AST-0204514 (for A.L.). The authors also
acknowledge the BSF grant 2004386.

\bsp

\label{lastpage}

\end{document}